\documentclass[twocolumn]{aastex63}

\usepackage[T1]{fontenc}
\usepackage[utf8]{inputenc}
\usepackage{xspace}
\usepackage{grffile}
\usepackage{multirow}
\usepackage{enumitem}

\graphicspath{{./}{figures/}}

\renewcommand{\arcsec}{\ensuremath{^{\prime\prime}}\xspace}
\renewcommand{\arcmin}{\ensuremath{^{\prime}}\xspace}

\newcommand{\kms}{km\,s$^{-1}$\xspace}

\newcommand{\sqcm}{cm$^{-2}$\xspace}
\newcommand{\cbcm}{cm$^{-3}$\xspace}

\newcommand{\Msun}{M$_\odot$\xspace}
\newcommand{\Msunyr}{M$_\odot$\,yr$^{-1}$\xspace}

\newcommand{\alphaUnits}{\Msun\,pc$^{-2}$\,(K\,\kms)$^{-1}$\xspace}
\newcommand{\co}[2]{CO(#1--#2)\xspace}

\newcommand{\astrodendro}{\textsc{astrodendro}\xspace}

\newcommand{\sigmaten}{\ensuremath{\sigma_\mathrm{10\,pc}}\xspace}

\newcommand{\Lten}{\ensuremath{L_\mathrm{10\,pc}}\xspace}
\newcommand{\avir}{\ensuremath{\alpha_\mathrm{vir}}\xspace}
\newcommand{\alphaCO}{\ensuremath{\alpha_\mathrm{CO}}\xspace}

\received{-}
\revised{-}
\accepted{20 July 2020}
\submitjournal{ApJ}

\shorttitle{Turbulent gas structure in NGC253 and the GC}
\shortauthors{Krieger et al.}

\begin{document}

\title{The turbulent gas structure in the centers of NGC253 and the Milky Way}

\correspondingauthor{Nico Krieger}
\email{krieger@mpia.de}

\author[0000-0003-1104-2014]{Nico Krieger}
    \affil{Max-Planck-Institut f\"ur Astronomie, K\"onigstuhl 17, 69120 Heidelberg, Germany}

\author{Alberto D. Bolatto}
    \affiliation{Department of Astronomy and Joint Space-Science Institute, University of Maryland, College Park, MD 20742, USA}
\author{Eric W. Koch}
    \affil{University of Alberta, Department of Physics, 4-183 CCIS, Edmonton AB T6G 2E1, Canada}
\author{Adam K. Leroy}
    \affiliation{Department of Astronomy, The Ohio State University, 4055 McPherson Laboratory, 140 West 18th Ave, Columbus, OH 43210, USA}
\author{Erik Rosolowsky}
    \affil{University of Alberta, Department of Physics, 4-183 CCIS, Edmonton AB T6G 2E1, Canada}
\author{Fabian Walter}
    \affil{Max-Planck-Institut f\"ur Astronomie, K\"onigstuhl 17, 69120 Heidelberg, Germany}
    \affiliation{National Radio Astronomy Observatory, P.O. Box O, 1003 Lopezville Road, Socorro, NM 87801, USA}
\author{Axel Wei\ss}
    \affiliation{Max-Planck-Institut f\"ur Radioastronomie, Auf dem H\"ugel 69, 53121 Bonn, Germany}

\author{David J. Eden}
    \affil{Astrophysics Research Institute, Liverpool John Moores University, IC2, Liverpool Science Park, 146 Brownlow Hill, Liverpool L3 5RF, UK}
\author{Rebecca C. Levy}
    \affiliation{Department of Astronomy, University of Maryland, College Park, MD 20742, USA}
\author{David S. Meier}
    \affiliation{New Mexico Institute of Mining and Technology, 801 Leroy Place, Socorro, NM 87801, USA}
    \affiliation{National Radio Astronomy Observatory, P.O. Box O, 1003 Lopezville Road, Socorro, NM 87801, USA}
\author{Elisabeth A. C. Mills}
    \affiliation{Department of Physics and Astronomy, University of Kansas, 1251 Wescoe Hall Dr., Lawrence, KS 66045, USA}
\author{Toby Moore}
    \affil{Astrophysics Research Institute, Liverpool John Moores University, IC2, Liverpool Science Park, 146 Brownlow Hill, Liverpool L3 5RF, UK}
\author{J\"urgen Ott}
    \affiliation{National Radio Astronomy Observatory, P.O. Box O, 1003 Lopezville Road, Socorro, NM 87801, USA}
\author{Yang Su}
    \affil{Purple Mountain Observatory and Key Laboratory of Radio Astronomy, Chinese Academy of Sciences, Nanjing 210034, China}
\author{Sylvain Veilleux}
    \affiliation{Department of Astronomy and Joint Space-Science Institute, University of Maryland, College Park, MD 20742, USA}

\begin{abstract}
We compare molecular gas properties in the starbursting center of NGC253 and the Milky Way Galactic Center (GC) on scales of $\sim1-100$\,pc using dendograms and resolution-, area- and noise-matched datasets in \co10 and \co32.
We find that the size -- line width relations in NGC253 and the GC have similar slope, but NGC253 has larger line widths by factors of $\sim2-3$. The $\sigma^2/R$ dependency on column density shows that, in the GC, on scales of $10-100$\,pc the kinematics of gas over $N>3\times10^{21}$\,\sqcm are compatible with gravitationally bound structures. In NGC253 this is only the case for column densities $N>3\times10^{22}$\,\sqcm. The increased line widths in NGC253 originate in the lower column density gas. This high-velocity dispersion, not gravitationally self-bound gas is likely in transient structures created by the combination of high average densities and feedback in the starburst. The high densities turns the gas molecular throughout the volume of the starburst, and the injection of energy and momentum by feedback significantly increases the velocity dispersion at a given spatial scale over what is observed in the GC.
\end{abstract}



\section{Introduction} \label{section: introduction}

The interstellar medium (ISM) in the centers of spiral galaxies differs in crucial ways from that in their disks. In strongly barred galaxies, the bar helps drive gas to the galaxy center \citep[e.g.][]{2019MNRAS.484.5192C}. This results in high gas surface densities, often organized into ring-like structures connected to the outer galaxy by dust lanes and gas streams \citep[e.g.][]{2001AJ....121..225B,2017MNRAS.471.4027B,2017MNRAS.470.3819B,2005A&A...429..141K,2013A&A...555L...4C,2011A&A...529A..45V}. The dynamical forces acting on the gas also differ in important ways, thanks to the deep potential well and nearly solid-body rotation often found in the centers of galaxies \citep[e.g.][]{2017MNRAS.466.1213K}. As a result, the gas contents of galactic centers are typically characterised by more extreme properties than the surrounding disk: higher densities, higher temperatures, and higher velocity dispersion and turbulence \citep[e.g.][]{Morris:1996db,2000ApJ...536..357M,2001ApJ...562..348O,2012MNRAS.425..720S,2013PASJ...65..118S,2016A&A...586A..50G,2017ApJ...850...77K,2019MNRAS.483.4291C,2019ApJ...871..170M}. 

Not all Galactic centers appear identical. Even among barred spiral galaxies, central regions vary dramatically in their gas content, star formation activity, and AGN activity \citep[e.g.,][]{kormendy2004}. Because achieving high physical resolution in other galaxies is challenging, it remains an open question how the detailed ISM structure varies along with the level of activity and gas mass in the centers of galaxies. Do more actively star-forming galaxies show increased turbulence? High surface densities at small scales?

In this paper we rigorously compare the parsec-scale molecular ISM structure between the Milky Way's relatively quiescent Galactic Center (GC) and the starbursting nuclear region in NGC253. After constructing carefully matched CO emission line datasets, we estimate the line width-size relation and surface density for each galaxy. We compare these to one another and the expectations for self-gravitating clouds. This is the first such high resolution, carefully matched comparison of molecular gas structure that we are aware of. Our work builds on previous studies of the GC line width -- size by \citet{2001ApJ...562..348O}, \citet{2012MNRAS.425..720S}, and a previous lower-resolution comparison of the two galaxy centers by \citet{Sakamoto:2011et}.

In Section~\ref{section: data}, we describe the datasets of NGC253 and the GC. We lay out our methods, describing dendrogroms and measurements in in Section~\ref{section: structural analysis}. We show the results (Section~\ref{section: results}), discuss them in Section~\ref{section: discussion} and summarise our work in Section~\ref{section: summary}. The appendix~\ref{appendix: structure definition} lists technical details and presents checks on our methods.

\subsection{The Quiescent Galactic Center}

The central $\sim 1$\,kpc of the Milky Way hosts $\approx 10\%$ of the total molecular gas mass of the Galaxy, with $\sim 6-8 \times 10^7$\,\Msun total molecular gas mass \citep[CMZ;][]{1998ApJS..118..455O,Morris:1996db,2007A&A...467..611F}. 

Despite this high gas mass and a high fraction of dense gas, the GC is often viewed as a relatively quiescent galaxy center. The integrated star formation rate (SFR) of the GC, $\sim 0.1$\,\Msunyr, is both lower than might be expected given the amount of dense gas \citep[e.g.,][]{2013MNRAS.429..987L,2017MNRAS.469.2263B} and low compared to starbursting galaxy nuclei like NGC253. This discrepancy has been attributed to cloud stabilization by dynamical effects or understood as catching the GC during a quiescent phase of an episodic or stochastic star formation history \citep[e.g.,][]{2015MNRAS.453..739K,2017MNRAS.466.1213K,2019MNRAS.484.1213S}. Evidence of winds and outflows from the GC hint towards more active star formation (or AGN activity) in the past \citep[e.g.,][]{2020ApJ...888...51L,2019MNRAS.482.4813S}.

We adopt the recent distance measurement of 8.178\,kpc for the GC \citep{2019A&A...625L..10G} for which 10\,pc correspond to 4.2\arcmin.
We refer to Sgr~A* at $l, b = 359.94422947^\circ, -0.04615714^\circ$ as the ``central position'' of the GC \citep{2011AJ....142...35P} and use 0\,\kms for the systemic velocity.

\subsection{The NGC253 Starburst}

The nearby galaxy NGC253 hosts the prototypical bar-fed nuclear starburst. Its central 500\,pc has a SFR $\sim 2$\,\Msunyr and a molecular gas reservoir of $\sim 3-4 \times 10^8$\,\Msun \citep{1996A&A...305..421M,2015ApJ...801...25L,2018ApJ...860...23P,2019ApJ...881...43K} fueled by gas accretion along the bar \citep{2004ApJ...611..835P}. 

This region hosts a collection of dense, massive molecular clumps \citep[e.g.][]{Sakamoto:2011et,2017ApJ...849...81A} that appear to be in the process of forming super star clusters \citep{2018ApJ...869..126L}.  The star formation drives a wind that has been observed in H$\alpha$, X-rays, and tracers of neutral and molecular gas \citep[e.g.][]{Turner:1985iy,Strickland:2000wd,2000ApJS..129..493H,Strickland:2002kp,2006ApJ...636..685S,Sharp:2010jl,Sturm:2011jb, Westmoquette:2011bp,2013Natur.499..450B,2017ApJ...835..265W,2019ApJ...881...43K}. 

Despite their differences, the spatial extent and orientation of NGC253 and the GC are quite similar. NGC253's inclination $i = 78^\circ$ compares well to the edge-on Milky Way GC. The two regions have similar size, $\sim 500$\,pc. And the physical resolution achieved by ALMA observations of NGC253 closely resemble that achieved by single dish mapping in the GC.

We adopt a distance to NGC 253 of 3.5\,Mpc \citep{Rekola:2005ha}, at which 10\,pc corresponds to 0.59\arcsec. We use the kinematic center at $\alpha, \delta = 00^h47^m33.134^s, -25^\circ17^m19.68^s$ \citet{MullerSanchez:2010dr} and adopt a systemic velocity of 250\,\kms .


\section{Data}
\label{section: data}

\begin{deluxetable*}{llccccll}
	\tablewidth{\linewidth}
	\tablecaption{Details of the datasets used in this analysis.}
	\label{table1}
	\tablehead{\colhead{set} & \colhead{source} & \colhead{line} & \multicolumn{2}{c}{resolution} & \colhead{noise\tablenotemark{b}} & \colhead{reference} & \colhead{ALMA ID}\\
	&&& \colhead{spectral} & \colhead{physical\tablenotemark{a}} &&&\\
	&&& \colhead{[\kms]} & \colhead{[pc]} & \colhead{[mK]} &
	}
	\startdata
	{\parbox[t]{2mm}{\multirow{2}{*}{\rotatebox[origin=c]{90}{low}}}} &
	  GC     & \co10 & 5.0 & 32.0 & 38 & \textsc{COGAL} \citet{2001ApJ...547..792D} &\\
	& NGC253 & \co10 & 5.0 & 32.0 & 38 & \citet{2013Natur.499..450B} & 2011.1.00172.S\\
	\rule{0pt}{4ex} 
	{\parbox[t]{2mm}{\multirow{2}{*}{\rotatebox[origin=c]{90}{high}}}} &
	  GC     & \co32 & 2.5 & \phantom{3}3.0 & 115 &  Eden et al., (in prep) &\\
	& NGC253 & \co32 & 2.5 & \phantom{3}3.0  & 115 & \citet{2019ApJ...881...43K} & 2015.1.00274.S\\
	\enddata
    \tablenotetext{a}{FHWM of the circular beam.}
    \tablenotetext{b}{Root mean square noise in line-free channels after matching the noise by adding beam-correlated Gaussian noise to the GC data.}
\end{deluxetable*}

We aim to compare the size -- line width relation, surface density, and dynamical state of the molecular gas between NGC253 and the GC at multiple scales. 

We trace the molecular gas using CO line emission. For a robust comparison requires us to compare the same tracers at the same physical resolution and sensitivity. Therefore, we construct matched CO datasets for the two galaxies, using \co10 for a low resolution comparison and \co32 for a high resolution comparison.

For NGC253, we use ALMA \co10 observations from \citet{2013Natur.499..450B}, \citet{2015ApJ...801...25L} and \citet{2015ApJ...801...63M} and ALMA \co32 from \citet{2019ApJ...881...43K}. The interferometric \co10 observations were carried out in ALMA cycle 2 and then combined with Mopra single dish observations. The final zero-spacing corrected data cube has 1.6\arcsec angular and 5.0\,\kms spectral resolution. The \co32 data were obtained with ALMA during cycle 4 and include total power observations. The resulting zero-spacing corrected data cube has 0.15\arcsec angular and 2.5\,\kms spectral resolution. More details regarding the data reduction can be found in the original publications.

We draw \co10  observations of the GC from the \textsc{COGAL} survey \citep{2001ApJ...547..792D}. 
These data have angular resolution 7.5\arcmin and spectral resolution 1.3\,\kms. Note that COGAL undersamples the Galactic plane at $\sim 1$ beam spacing, and data have been interpolated to obtain a filled map \citep{2001ApJ...547..792D}.
For \co32, we use observations of the GC obtained by the \textsc{CHIMPS2} project (Eden et al., in prep.), which extends the \textsc{CHIMPS} Galactic plane survey \citep{2016MNRAS.456.2885R} into the the inner galaxy.  The data build on the data reduction recipe of COHRS \citep[CO high-resolution survey of the Galactic plane;][]{2013ApJS..209....8D}.
\textsc{CHIMPS2} achieves 15.0\arcsec spatial and 1.0\,\kms spectral resolution.

We match the data between the two galaxies as closely as possible, constructing data cubes with identical spatial and spectral resolution, pixel scale, orientation with respect to the galactic plane, field of view (FoV), and noise. The following steps were followed:

\begin{enumerate}[noitemsep,topsep=0pt,leftmargin=2\parindent]
\item[(1)] The images are smoothed to circular beams with the highest possible common resolution (32\,pc for \co10 and 3.0\,pc for \co32).
\item[(2)] The images are then reprojected onto a common pixel grid aligned with galactic longitude and latitude. In NGC253 we defined the galactic plane to lie along the major axis of the galaxy. We used pixel scales of 6.4\,pc and 0.6\,pc for \co10 and \co32, oversampling the beam by a factor of 5.
\item[(3)] The spectral resolution is matched at 5.0\,\kms for \co10 and 2.5\,\kms for \co32. The data were reprojected onto a matched velocity grid covering from $-250$\,\kms to $+250$\,\kms about the systemic velocity for both sources and both lines.
\item[(4)] The field of view is restricted to the overlap between the images so that we study the same amount of area in each galaxy. Centered on the respective galactic center, the FoVs are 1500\,pc by 750\,pc for the wider FoV in \co10 and 800\,pc by 400\,pc in \co32. 
\item[(5)] After these steps, the noise in the datasets varies by a factor of $\sim 2$ between the NGC253 and the GC datasets. To keep the analysis consistent between the two galaxies, we add additional beam-correlated\footnote{Random noise that has been convolved with a Gaussian beam and scaled to the appropriate level.} Gaussian noise to both GC images to match them to the higher noise of the NGC253 observations. For the CHIMPS2 data, the noise varies spatially across the map and we add noise as needed to achieve a uniform noise level. The final rms noise is 38\,mK in a 5.0\,\kms channel in \co10 and 115\,mK in a 2.5\,\kms channel in \co32.
\end{enumerate}

The final image parameters are given in Table~\ref{table1}.


\section{Methods}
\label{section: structural analysis}

We use dendrograms to identify CO-emitting structures at multiple scales and then measure their line width, luminosity, and size. Using these measurements, we compare the line width and surface density between the two systems at many spatial scales.

We are particularly interested in the size -- line width relation and the relationship between line width, size, and surface density, which traces the dynamical state of the gas. We then look for ways in which the different overall gas mass and level of star formation activity in the GC and NGC253 may affect the gas structure.

\subsection{Dendrogram structure identification}
\label{section: dendrogram}

We use dendrograms to identify distinct CO-emitting structures at multiple scales. Detailed descriptions of this method are given by 
\citet[][]{2008ApJ...679.1338R}, \citet{Goodman:2009dp}, \citet{2012MNRAS.425..720S}, and on the \astrodendro homepage\footnote{\url{dendrograms.readthedocs.io}}. Briefly, the algorithm identifies structures using a series of iso-intensity contours. As the contour level drops, individual discrete ``leaves'' identified at the highest contours merge into ``branches,'' which combine multiple substructures. Eventually these larger structures merge together into a ``trunk,'' which will not merge with any other structures.
Since the structure identification is hierarchical, a given voxel in the original data cube can be included in several nested structures.

The hierarchical nature of the dendrogram approach is ideal to extract multi-scale information from our high spatial dynamic range data. Recently, \citet{2020RAA....20...31L} tested several common clump detection algorithms and found so-called dendrograms to be among the best methods due to high accuracy and detection completeness.

We use the \astrodendro implementation, which has several tuning parameters. We only consider emission with SNR $>5$ (cf. Table~\ref{table1}). The minimum difference in intensity between nested structures is set to one times the rms noise.
We require the minimum phase space volume per structure to be three times the spatial resolution element times velocity channel width. The choices ensure that we focus on significant, well-resolve structures.
We further restrict our analysis to the scales on which we can resolve the structures but large-scale motions do not yet dominate the observed motions. We list these in Table~\ref{table2}. 

\begin{deluxetable}{llcc}
    \tablewidth{\textwidth}
    \tablecaption{Limits on recoverable structure sizes.}
    \label{table2}
    \tablehead{\colhead{source} & \colhead{line} & $R_\mathrm{min}$ & $R_\mathrm{max}$\\
    & & [pc] & [pc]\\
    & & (1) & (2)
    }
    \startdata
        NGC253 & \co10 & 6.0 & 72\\
        GC     & \co10 & 8.5 & 79\\
        NGC253 & \co32 & 0.55 & 18\\
        GC     & \co32 & 0.90 & 20\\
    \enddata
    \tablecomments{(1) Completeness limit imposed by the minimum-volume-of-a-structure threshold.\\
    (2) Limit beyond which large scale dynamics (e.g. galactic rotation) dominate structure properties.}
\end{deluxetable}

\subsection{Measured quantities}
\label{section: measure}

The dendrograms identify $\approx 24,000$ position-position-velocity structures of interest across our data. For each structure, we measure the size, line width, and luminosity and calculate the implied mass and column density:

\subsubsection{Size}
\label{section: dendrogram size}

We define the size, $R$, of a structure as the geometric mean of the semi-major and semi-minor axis size. We compute these using the 
intensity-weighted second moment, with the major axis defined as the direction of greatest elongation of the object. This definition is implemented in \astrodendro as the \texttt{radius} quantity.

Note that for a Gaussian cloud, this definition of size corresponds to the $1\sigma$ value, while we quote beam sizes as FWHM. Also note that we do not deconvolve the beam from the structure size, trusting that our minimum volume requirement leads us to select only well-resolved structures.

We confirm with tests that this definition of size do not affect our analyses as other definitions merely shift the normalization of the sizes (cf. Appendix~\ref{appendix: size definition}).

\subsubsection{Line width}
\label{section: dendrogram line width}

We define line width, $\sigma$, as the the intensity-weighted second moment, i.e., the intensity-weighted velocity dispersion, over all pixels belonging to the structure. This is implemented in \astrodendro as the \texttt{v\_rms} quantity. Note that we do not deconvolve the channel width from the measured line widths. We also make no correction for galactic rotation or other bulk flows, which can contaminate the measurements on large scales.

Analog to size, we confirm with tests that other definitions of line width do not affect our analyses beyond shifting the line width normalization (cf. Appendix~\ref{appendix: line width definition}).

\subsubsection{Luminosity}
\label{section: luminosity}

We calculate the luminosity of each structure as the area- and line-integrated intensity $L = \int I \mathrm{d}A$, where $A$ refers to the spatial area and $I$ to the line integrated intensity $I = \int I_\nu dv$. The integrated intensity $\sum_i I_i$ is reported by \astrodendro . We apply channel width and pixel area corrections to derive the luminosity.

\subsubsection{Mass}
\label{section: dendrogram mass}

From the luminosity, we estimate the molecular gas mass of each structure. For \co10 , we do this via $M = \alphaCO L$, where \alphaCO is the CO(1-0)-to-H$_2$ conversion factor \citep[see][]{2013ARAA..51..207B}. For \co32  we calculate $M = \alphaCO r_{31} L$, where $r_{31}$ is the empirical line ratio to translate from \co32 to \co10 luminosity.

The exact \alphaCO for galactic centers remains uncertain, though it certainly appears  lower than the standard Milky Way disk value \citep[e.g.,][]{2013ARAA..51..207B}. We adopt $\alphaCO = 2.2$\,\alphaUnits, i.e., half the nominal Solar Neighborhood value, for the Galactic Center \citep[see discussion in][]{2013ARA&A..51..511K}. We adopt a lower value $\alphaCO = 1.1$\,\alphaUnits, i.e., one quarter the Solar Neighborhood value, for NGC253. This lower value in NGC253 is partially motivated by observations \citep[e.g.,][]{2015ApJ...801...25L} and partially by the expectation that the denser, excited gas in NGC253 \citep[e.g.,][]{2019ApJ...871..170M} should show starburst-like conversion factors. These numbers do not affect two of our three main results, the size -- line width relation or the size -- luminosity relation. The value of \alphaCO does affect the estimate dynamical state of the gas. We return to the impact of our assumption when discussing this below.

For the line ratio, we adopt $r_{31} = I_{3-2}\,I_{1-0} ^{-1} = 0.67$ for both galaxies to keep the analysis consistent. Across each source, we measure $r_{31} = 0.63$ in NGC253 and $r_{31} = 0.68$ in the GC after matching the area between the \co32 and \co10 maps. For the dendrogram structures, $r_{31}$ may deviate from the global average and variations in $r_{31}$ will linearly scale the \co32 -based masses.

\subsubsection{Column density}
\label{section: dendrogram column density}

We also estimate the average column density of each structure. To do this, we divide the mass by an luminosity-weighted elliptical area, $A_\mathrm{eff}$, calculated from the major and minor axis as described above. Then we convert to units of H$_2$ molecules per cm$^{-2}$. This column density N$_{\rm H2}$ does not include helium.


\subsection{Binned analysis}
\label{section: binning}

The dendrogram analysis yields many structures, e.g. $>12,000$ the \co32 data of NGC253 alone. Here we primarily focus on the average properties of structures at a given size scale or surface density. We bin the properties measured for individual structures to access these average properties. We create two sets of bins. First we bin alls structures by their size, $R$, in 0.1\,dex-wide bins. In each of these bins, we measure the median and $16^{th}$ percentile to $84^{th}$ percentile range of the line width and mass. Second, we bin the structures by surface density using bins 0.25\,dex wide. In each of these bins, we measure the median and  $16^{th}$ percentile to $84^{th}$ percentile range of the line width, size, and the size -- line width coefficient, $\sigma^2/R$, which we use to assess the dynamical state of the gas below.

At the low end, the minimum volume of a structure limits the size measurements ($R_\mathrm{min}$ in Table~\ref{table2}). This limit appears as a diagonal cutoff in size-line width space. Note that the GC shows lower line width at fixed size than NGC253, as a result the minimum-volume threshold imposes a higher minimum size in the GC. Also note that this effectively excludes the small structures that are most affected by beam convolution and channel convolution. Even in our smallest bins, beam deconvolution effects are $< 25$\% and because this has no effect on the analysis, we do not include any correction.

We identify the upper end of the analysis range ($R_\mathrm{max}$ in Table~\ref{table2}) as the size scale at which the measured line width jumps to very high values. This occurs when galactic motions dominate. The transition to this regime is sharp, making it easy to identify an upper size limit by eye.


\subsection{Fitting}
\label{section: fitting}

We conduct power law fits to the binned data. We fit $\sigma$ as a function of $R$ (``the size -- line width relation'') 

\begin{equation}
\label{equation: size-line width}
\sigma = a R^{b}
\end{equation}

\noindent and $L$ as a function of $R$ (the ``size -- luminosity relation''):

\begin{equation}
    L = c R^{d}
    \label{equation: size-luminosity}~.
\end{equation}

\noindent To do this we fit the bin centers and median values as lines in log log space using a weighted least squares minimization. We adopt the square root of the diagonals of the covariance matrix as the uncertainties, but note that these statistical errors are often small and systematic errors are non-negligible. We discuss this more below. In addition to reporting the exponents $b$ and $d$, we report the coefficients normalized to an intermediate size scale in our data, $R=10$\,pc, \sigmaten and \Lten .


\section{Results}
\label{section: results}

We apply the dendrogram analysis to both lines in both galaxies. Details of the dendrogram statistics and power law fits to the size -- line width relation and size -- luminosity relation are listed in Table~\ref{table3}. The binned data is available in the machine-readable format and a preview is given in Table~\ref{table4}.
In the following, we present the derived size -- line width and size -- luminosity relations. For each analysis, we first present data on the GC and NGC253 followed by a comparison. In the next section, we connect these measurements via an analysis of the size -- line width coefficient.

\begin{deluxetable*}{lcrrrcccc}
        \tablewidth{\textwidth}
        \tablecaption{Dendrogram statistics and fit results for power law fits to the binned size -- line width and size -- luminosity relations shown in Figures~\ref{fig2} and \ref{fig3}.}
        \label{table3}
        \tablehead{\colhead{source} & \colhead{line} & \multicolumn{3}{c}{dendrogram structures} & \multicolumn{2}{c}{size -- line width} & \multicolumn{2}{c}{size -- luminosity}\\
        && total & branches & leaves & $b$ & \sigmaten & $d$ & log $\Lten$\\
    &&&&& (1) & (2) & (3) & (4)
        }
        \startdata
NGC253 & CO(1-0) &   991 &  466 &  520 & $0.82\pm0.02$ & $\phantom{1} 8.9\pm0.2$ & $2.92\pm0.07$ & $4.27\pm0.11$ \\
GC     & CO(1-0) &   324 &  158 &  165 & $0.74\pm0.04$ & $\phantom{1} 3.3\pm0.4$ & $3.25\pm0.13$ & $4.34\pm0.20$ \\
NGC253 & CO(3-2) & 12414 & 5145 & 7024 & $0.62\pm0.01$ & $17.1\pm0.1$ & $2.89\pm0.02$ & $5.44\pm0.03$ \\
GC     & CO(3-2) & 10235 & 4563 & 5570 & $0.72\pm0.03$ & $\phantom{1} 8.9\pm0.2$ & $2.69\pm0.02$ & $4.96\pm0.02$ \\
    \enddata
    \tablecomments{The errors are formal errors, which assume independent, Gaussian distributed data, and underestimate the range of slopes that could be accommodated in Figures~\ref{fig2} and \ref{fig3}.\\
                   (1) Exponent $b$ of the power law fit to the size -- line width relation according to Equation~\ref{equation: size-line width}.
                   (2) Characteristic line width at 10\,pc according to the power law fit to the size -- line width relation (Equation~\ref{equation: size-line width}) in \kms.
                   (3) Exponent $d$ of the power law fit to the size -- luminosity relation according to Equation~\ref{equation: size-luminosity}.
                   (4) Characteristic luminosity at 10\,pc according to Equation~\ref{equation: size-luminosity} in $\log \mathrm{M}_\odot$.}
\end{deluxetable*}

\begin{deluxetable*}{llccccc}
        \tablewidth{\linewidth}
        \tablecaption{Sample of the binned size -- line width data for \co32 in NGC253. All data of the size -- line width, size -- luminosity and column density -- $\sigma^2$/R relations for both tracers and both sources is available in the machine-readable format in the online journal. The table shown here provides guidance regarding the form and content.}
        \label{table4}
        \tablehead{\colhead{galaxy} & \colhead{CO} & \colhead{R$_\mathrm{min}$} & \colhead{R$_\mathrm{max}$} & \colhead{$\sigma_\mathrm{16th}$} & \colhead{$\sigma_\mathrm{median}$} & \colhead{$\sigma_\mathrm{84th}$} \\
        && [pc] & [pc] & [km\,s$^{-1}$] & [km\,s$^{-1}$] & [km\,s$^{-1}$]\\ \relax
    && (1) & (2) & (3) & (4) & (5)
        }
        \startdata
NGC253 & CO(3-2) &  0.49 &  0.62 &  2.24 &  2.87 &  4.10\\
NGC253 & CO(3-2) &  0.62 &  0.78 &  2.22 &  3.17 &  4.65\\
NGC253 & CO(3-2) &  0.78 &  0.98 &  2.53 &  3.71 &  5.66\\
NGC253 & CO(3-2) &  0.98 &  1.23 &  3.00 &  4.35 &  6.81\\
NGC253 & CO(3-2) &  1.23 &  1.55 &  3.35 &  4.99 &  7.86\\
NGC253 & CO(3-2) &  1.55 &  1.95 &  3.93 &  5.76 &  9.13\\
NGC253 & CO(3-2) &  1.95 &  2.46 &  4.42 &  6.69 & 10.73\\
NGC253 & CO(3-2) &  2.46 &  3.09 &  4.94 &  7.43 & 11.69\\
NGC253 & CO(3-2) &  3.09 &  3.89 &  6.24 &  9.31 & 14.06\\
NGC253 & CO(3-2) &  3.89 &  4.90 &  7.00 & 10.70 & 16.75\\
NGC253 & CO(3-2) &  4.90 &  6.17 &  8.14 & 13.33 & 20.37\\
NGC253 & CO(3-2) &  6.17 &  7.77 &  9.16 & 11.94 & 23.16\\
NGC253 & CO(3-2) &  7.77 &  9.78 &  9.42 & 17.68 & 27.99\\
NGC253 & CO(3-2) &  9.78 & 12.31 & 12.75 & 21.13 & 26.05\\
NGC253 & CO(3-2) & 12.31 & 15.50 & 14.00 & 23.03 & 30.71\\
    \enddata
    \tablecomments{(1) Lower edge of the size bin.
                   (2) Upper edge of the size bin.
                   (3) Lower bound of the line width distribution (16th percentile).
                   (4) Median of the line width distribution (50th percentile).
                   (5) Upper bound of the line width distribution (84th percentile).
                  }
\end{deluxetable*}


\subsection{Size -- line width relation}
\label{section: size line width}

\begin{figure*}
    \centering
    \includegraphics[width=0.8\textwidth]{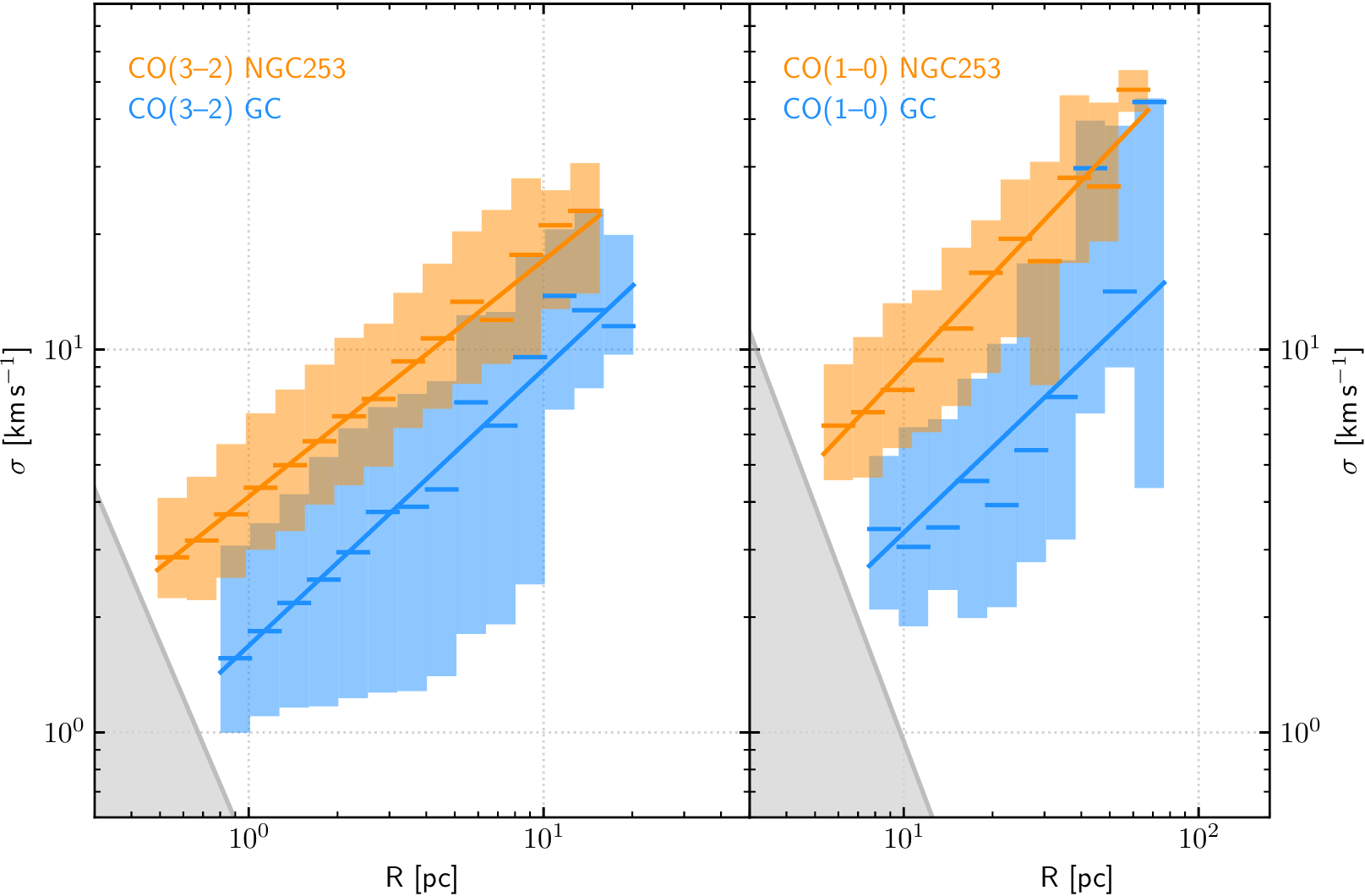}
    \caption{Binned size -- line width relation for \co32 (\emph{left}) and \co10 (\emph{right}) in NGC253 (orange) and the GC (blue). Horizontal lines indicate the median line width in each bin. The shaded colored region indicates the $16^{th}$ to $84^{th}$ percentile range of line widths in that size bin. The values of the power law fits (solid lines) are given in Table~\ref{table3}. 
    The grey, shaded areas show regions where no structures could be detected because of our minimum volume limit. The size -- line width relation in NGC253 is significantly offset towards larger line widths from the relation in the GC for both tracers.
    \label{fig2}}
\end{figure*}

Figure~\ref{fig2} presents the binned size -- line width relations for the GC and NGC253. With the high resolution \co32 data, we are able to cover the size range down to $<1$\,pc whereas the lower resolution \co10 covers the larger scales up to $\sim 80$\,pc. As discussed above, we omit the largest size scales, on which we expect galactic rotation and large scale motions to contaminate the line width.

These measurements capture the hierarchical structure of the input data. Given this, a given structure does not necessarily correspond to a (giant) molecular cloud. Especially for the \co32 data, the small leaves on top of nested branches are likely not independent bound clouds, but represent substructure that could be described as ``cloudlets'' or ``cores'' within a cloud. Structures at larger scales may represent associations of multiple bound structures.

\subsubsection{Galactic Center}
\label{section: size line width: GC}

The median trend in the data is reasonably well represented by a power law fit of the form in Eq.~\ref{equation: size-line width}, although we do measure considerable scatter about the median. The fitted slopes are $b=0.72 \pm 0.03$ in \co32 and $b=0.74 \pm 0.04$ in \co10. 
The typical line width at 10\,pc derived from the fit is $\sigmaten = 8.9 \pm 0.2$\,\kms in \co32 and $\sigmaten = 3.3\pm0.9$\,\kms in \co10.


\subsubsection{NGC253}
\label{section: size line width: NGC253}

In NGC253, the median of the binned \co10 and \co32 data (Figure~\ref{fig2}) almost perfectly follows a power law over more than one order of magnitude. A fit results in exponents of $b=0.62 \pm 0.01$ in \co32 and $b=0.82 \pm 0.02$ in \co10. 
The fit yields typical line widths of $\sigmaten = 17.1 \pm 0.1$\,\kms in \co32 and $\sigmaten = 8.9 \pm 0.2$\,\kms in \co10.


\subsubsection{Comparison of the size -- line width relations}
\label{section: size line width: comparison}

The size -- line width relations in the GC and NGC253 have similar slopes but are significantly offset in normalization. When parametrized by Eq.~\ref{equation: size-line width}, the line widths are wider in NGC253 by a factor of $\sim 1.9$ for \co32  and $\sim 2.7$ for \co10 .

The GC shows a wider distribution of line widths than NGC253 at a fixed size scale, as demonstrated by the larger vertical color bars in Figure~\ref{fig2}. This suggests a greater variation of cloud properties in the GC compared to more uniform structures in NGC253.

In both galaxies, the \co32 line widths appear broader than the  \co10 line widths. At overlapping scales ($8-16$\,pc), the \co32  line widths appear $\sim 1.9$ times broader than \co10  in NGC253 and $\sim2.7$ times broader in the GC. 

The lower line widths in \co10 appears to partially result from the poorer resolution of those data. As a test, we degrade the resolution of the \co32 data in NGC253 and repeat the dendrogram analysis. Degrading the spatial resolution (6.4\,pc to 32\,pc in ten steps to match the \co10 data) and the spectral resolution (2.5\,\kms to 5.0\,\kms) causes a shift of the size -- line width relation towards larger sizes (to the right-hand side in Figure~\ref{fig2}) that is approximately linear with resolution. At a fixed size scale, the measured line width is thus smaller with lower resolution data. Half of the observed line width mismatch between \co32  and \co10  can be explained directly as a consequence of these resolution effects. 

The other half likely arises from the fact that \co32  traces denser gas, usually associated with higher surface densities and more massive structures which correspondingly have larger line widths.


\subsection{Size -- luminosity relation}
\label{section: size luminosity}

\begin{figure*}
    \centering
    \includegraphics[width=\textwidth]{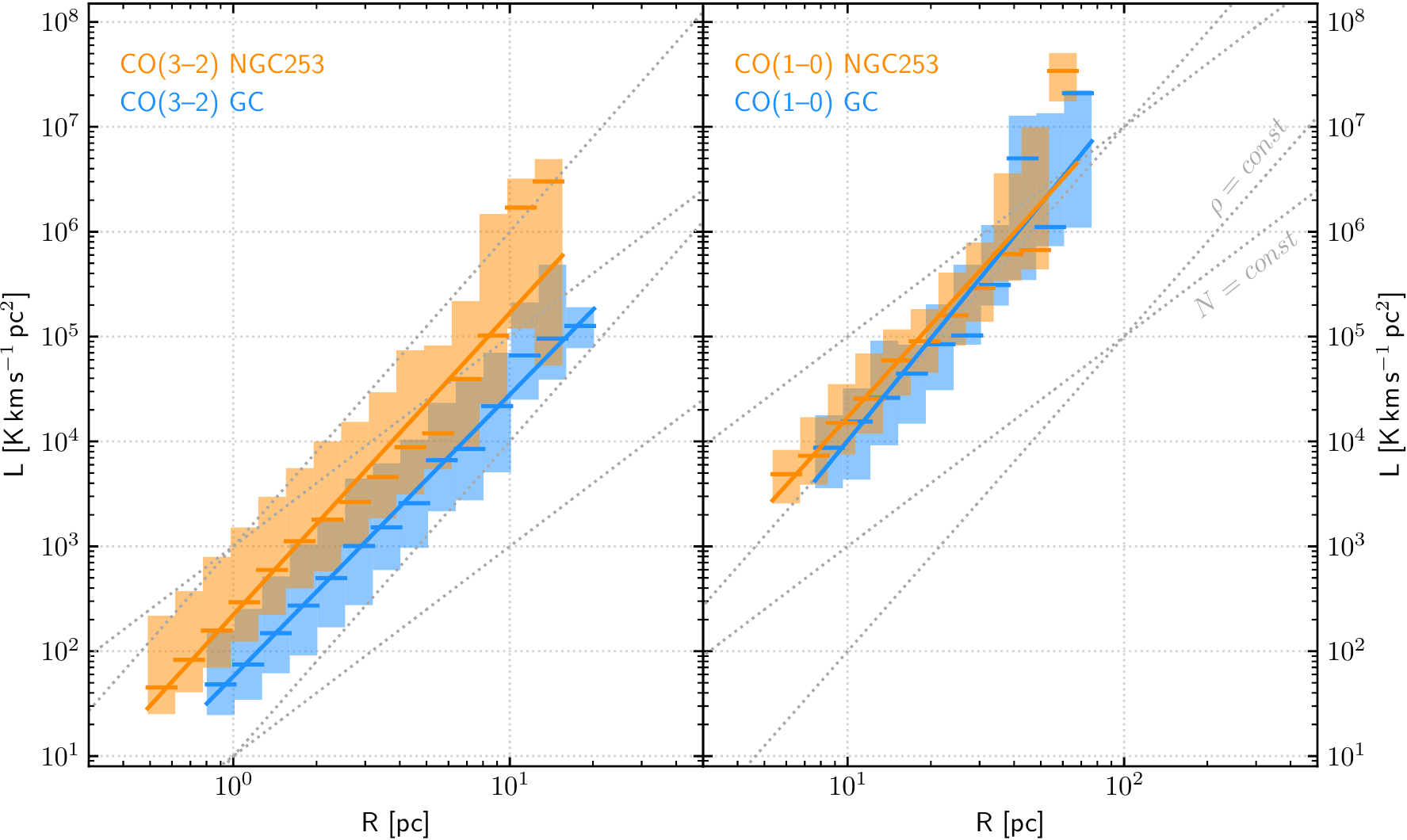}
    \caption{Relation between dendrogram structure size and luminosity for \co32 (\emph{left}) and \co10 (\emph{right}) in NGC253 and the GC. Horizontal lines indicate the median luminosity in each bin. Shaded regions show the 16$^{\rm th}$ to 84$^{\rm th}$ percentile range. The values of the power law fits (solid lines) are given in Table~\ref{table3}. In each panel, dotted lines illustrate two lines of constant column density $N$ ($L \propto M \propto R^2$) and constant volume density $\rho$ ($L \propto M \propto R^3$), calculated assuming fixed \alphaCO.
    }
    \label{fig3}
\end{figure*}

Figure~\ref{fig3} shows the size -- luminosity relation and Table~\ref{table3} lists the power law fit parameters according to Equation~\ref{equation: size-luminosity}.


\subsubsection{Galactic Center}
\label{section: size luminosity: GC}

The binned size -- luminosity relations for \co10 and \co32 in the GC are well represented by power laws.
The fits yield power law exponents of $d=2.69 \pm 0.02$ in \co32 and $d=3.25 \pm 0.13$ in \co10.  Given that the formal error bars understate the uncertainty, as the data are correlated across $R$, and if the largest bin is discarded, the \co10 exponent is consistent with $d=3$.

This is steeper than the $d=2$ that would indicate fixed surface brightness structures, so that larger structures show higher surface brightness. The fitted slope is more similar to the $d=3$ expected for structures with fixed volume density, assuming a constant \alphaCO. However, \alphaCO may well change as a function of luminosity or scale \citep[e.g., as found by][]{1987ApJ...319..730S} and this will also affect the slope of the size -- luminosity relation.


\subsubsection{NGC253}
\label{section: size luminosity: NGC253}

In NGC253, the binned size -- luminosity relations also scale nearly perfectly as power laws with exponents of $d=2.89 \pm 0.02$ in \co32 and $d=2.92 \pm 0.07$ in \co10. As in the GC, these exponents are closer to the $d \sim 3$ expected for constant volume density and \alphaCO than the $d \sim 2$ expected for constant surface brightness.


\subsubsection{Comparison of the size -- luminosity relation}
\label{section: size luminosity: comparison}

Figure~\ref{fig3} shows broad similarities between the size -- luminosity relations in NGC253 and the GC. Both galaxies exhibit slopes in the range $d = 2.7{-}3.3$ for both lines. The \co32 size -- luminosity relations are offset by a factor $\sim 2-3$ with higher luminosities in NGC253. In \co10, the relations almost perfectly overlap on scales of $\sim 8{-}50$\,pc. We do caution that this applies only to the range of plotted scales: on 1\,kpc scales the \co10 luminosity of the NGC253 nucleus is $3-4$ times higher than that of the GC \citep{1996ApJ...456L..91J}. Some of this separation is already visible in the largest-scale \co10 bin. 

Although the integrated line ratios are similar, more bright \co32 substructure might be expected in NGC253 compared to the GC in \co32 because of the more intense ongoing star formation activity in that galaxy. Reflecting this activity, the excitation of NGC253 has been measured to be higher \citep[e.g.,][]{bradford2003,2019ApJ...871..170M}.

As above, we emphasize that Figure \ref{fig3} reflects a measurement of hierarchical structure. The plots show that dendrogram-extracted substructures with matched sizes have similar \co10  luminosities in the two galaxies, not that the overall \co10 distribution has the same distribution or overall luminosity in the two cases.

For reference, we also calculate the size -- mass relations by applying our adopted CO-to-H$_2$ conversion factors (cf. Section~\ref{section: dendrogram mass}) in Appendix~\ref{appendix: size mass}.


\section{Discussion}
\label{section: discussion}


\subsection{Virial state of the molecular gas}
\label{section: virial state}

\begin{figure*}
    \centering
    \includegraphics[width=\textwidth]{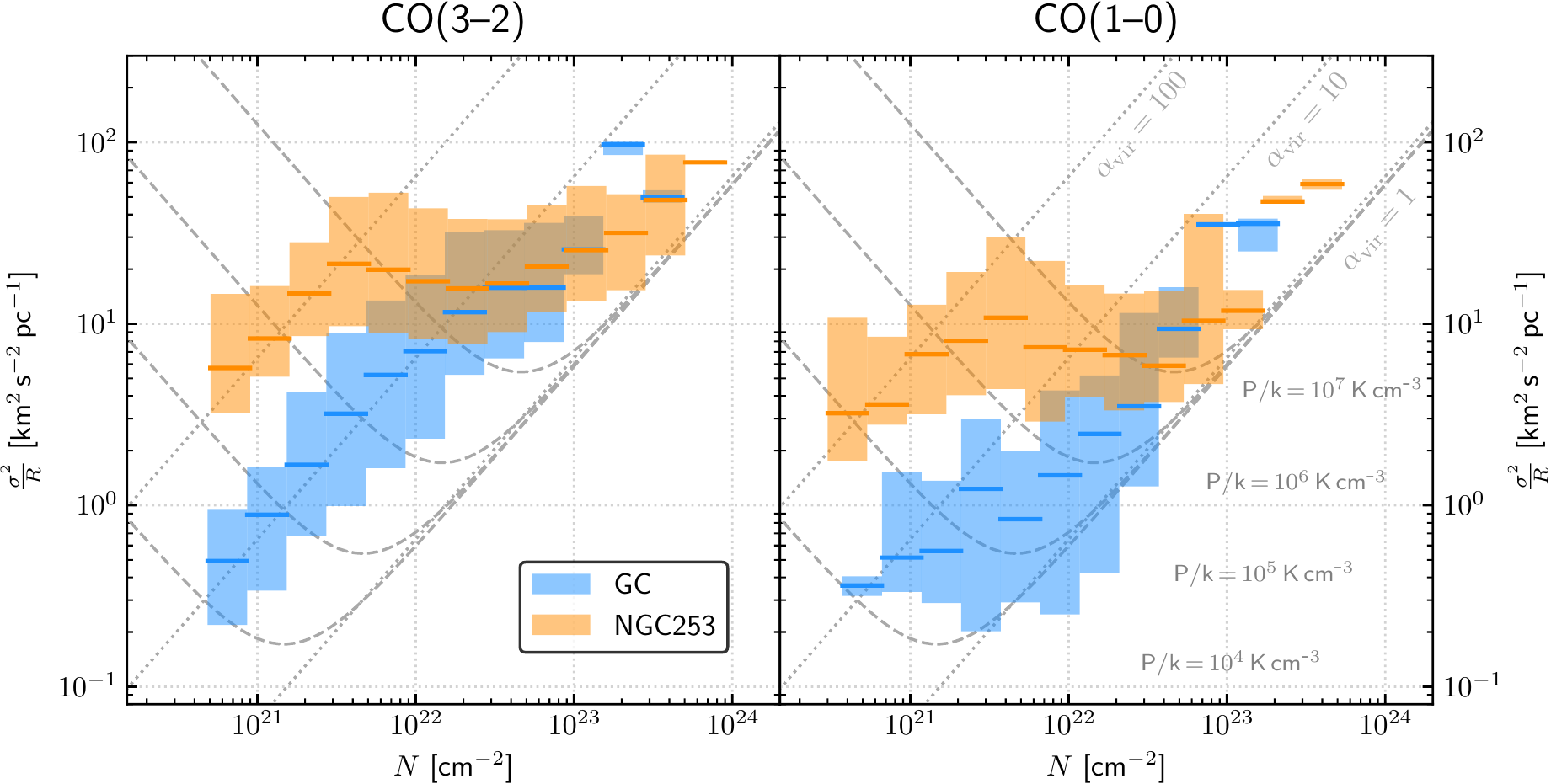}
    \caption{Size--line width coefficient as a function of column density in NGC253 and the GC under the assumption that luminous (CO detected) mass traces virial (gravitational) mass. Diagonal lines indicate lines of constant virial parameter under the assumption of idealized spherical clouds (cf. Section~\ref{section: virial state}). Dashed lines represent lines of constant external pressure on a spherical cloud \citep[][see Section~\ref{section: virial state}]{2011MNRAS.416..710F}. Horizontal lines indicate the median of the distribution of $\sigma^2/R$ (colored bars) in each bin.\\
    \emph{Left}: high resolution \co32; \emph{Right}: low resolution \co10.
    Note that the derived column density should be considered a lower limit (cf. Section~\ref{section: physical implications}). A different choice of conversion factors shifts the obtained relations along the x-axis but does not influence the slope.
    Due to the similar geometry and gas distribution in NGC253 and the GC, a relative comparison is still possible even if the absolute values must be interpreted with care. The strongly enhanced $\sigma^2/R$ at column densities $N \lesssim 3 \times 10^{22}$\,\sqcm implies that the low column density molecular gas in NGC253 is gravitationally unbound which is not the case in the GC. Appendix~\ref{appendix: virial state separated} shows this plot separated in size bins to address the degeneracy between $\sigma$ and $R$ in the size -- line width coefficient.
    }
    \label{fig4}
\end{figure*}

Substructures in NGC253 show higher line width at fixed size scale compared to the GC. In this section, we compare the measured line widths at fixed size to expectations for gravitationally bound clouds in virial equilibrium to explore the origins of the observed line widths.

Massive molecular clouds in the disks of the Milky Way and nearby galaxies are typically found to be close to being gravitationally bound (e.g., clouds in the Galactic disk, \citealt{1987ApJ...319..730S} and \citealt{2006ApJS..163..145J}; M51, \citealt{2014ApJ...784....3C}; NGC300, \citealt{2018ApJ...857...19F}; and for a large recent synthetic analysis see \citealt{2018ApJ...860..172S}).

Deviation from virial equilibrium is often expressed as the virial parameter $\avir = 2K / U$ where $K$ is the kinetic energy and $U$ the gravitational potential. Following \citet{2018ApJ...860..172S}, who follow \citet{keto86}  and \citet{Heyer:2009ii}, for idealized spherical clouds the line width $\sigma$, virial parameter \avir, size $R$ and average column density $N$ relate via
\begin{equation}
\label{eq:virial}
    N = \frac{5}{f \avir G \pi} \frac{\sigma^2}{R},
\end{equation}

\noindent where $G$ is the gravitational constant, the factor $f = (1-\gamma/3)/(1-2\gamma/5)$ accounts for the internal cloud structure with a radial density profile $\rho(r) \propto r^{-\gamma}$. For an isothermal cloud with $\gamma = 2$, thus $f = 5/3$. In Equation \ref{eq:virial}, the square of the coefficient of the size -- line width relation, $\sigma^2/R$, depends directly on the column density of the cloud, $N$. This idealized case is a vast oversimplification for real molecular clouds but the deviation from this case can yield insight into its dynamical state and the relative contribution of gravity and forces such as external pressure or magnetic support to the measured line width.

External pressure will broaden the observed line width and increase $\sigma^2/R$. Under the assumption of virial equilibrium, the effect of external pressure on  $\sigma^2/R$ is described by \citet{2011MNRAS.416..710F} as

\begin{equation}
\label{eq:press}
    \frac{\sigma^2}{R} = \frac{1}{3} \left( \pi \Gamma G \Sigma + \frac{4P_e}{\Sigma}\right)
\end{equation}

\noindent where $G$ is the gravitational constant, $\Sigma$ the mass surface density and $P_e$ the external pressure. 
$\Gamma$ is a form factor of order unity \citep{1989ApJ...338..178E} and we here use $\Gamma=0.73$ for an isothermal spherical cloud of critical mass.
We include the contribution of helium to the mass and derive the mass surface density $\Sigma_{mol} = 1.36 \times 2\,\mathrm{u}\times N = 2.16\times10^{-20} N$ from the column density $N$ in cm$^{-2}$ to units of M$_\odot$\,pc$^{-2}$.

Figure~\ref{fig4} plots $\sigma^2/R$ as a function of $N$ for both galaxies and and both lines. The dotted diagonal lines show fixed \avir\ in the absence of external pressure, from Equation \ref{eq:virial}. The curved dashed lines show virial equilibrium for different external pressures (Equation \ref{eq:press}). 
Objects that lie far above the $\avir = 1$ line may either be in equilibrium with a substantial pressure exerted by an external medium or they may be transient and/or out of equilibrium with an excess of kinetic energy over gravitational energy. Structures that are small in size and/or low mass are, generally speaking, less likely to be in equilibrium with self-gravity \citep{Heyer:2001}. 

Our results for in Figure~\ref{fig4} show a combination of both self-gravitating and high $\avir$ substructure. Recall that the \co10  data sample structures with sizes $\sim10-100$\,pc. In the GC the structures picked out in the \co10  analysis seem to approximately follow the expectations for gravitational bound objects with virial parameters $\avir \sim 2-3$. Thus \co10  in the GC, on average on 10-100\,pc, the line widths recovered by the dendrograms are consistent with those expected from self gravity and marginally bound, $\avir=2$, gas.

In the \co10  measurements for the center of NGC253 two regimes are apparent, splitting at a column density $N\sim3\times10^{22}$\,cm$^{-2}$. At lower column densities $\sigma^2/R$ is approximately constant at $\sim8$~km$^2$s$^{-2}$pc$^{-1}$, while for larger columns the trend appears likely to be the same as for the GC. At high column densities, our results agree with the measured line widths and dynamical state obtained by \citet{2015ApJ...801...25L} studying 20-30\,pc sized GMC-like structures. At lower column densities, the structures are likely either transient or in equilibrium with a high external pressure. Both of these possibilities are likely, although we consider it more likely that our measurements are dominated by transient structures found by the dendrogram decomposition. 

The preponderance of transient structure appears even more marked in \co32, which mostly samples scales of $1-10$\,pc. On these smaller scales the substructure found by the dendograms in both galaxies only appear to be self-gravitating at the largest column densities. 

We know that some very massive, self-gravitating structures exist on these small scales in NGC253, we have identified molecular clumps associated with young, massive clusters \citep{2018ApJ...869..126L}. These are certainly held together by gravity, though the stars may contribute to the potential. We also know that these structures are not particularly prominent in the CO~(3-2) maps \citep{Krieger2020}, which show bright CO emission throughout the starburst. Massive structures on 1-10\,pc scales are also known in the GC. Several of the GMCs identified by \citet{2001ApJ...562..348O} in \co10  have sizes $\lesssim 10$\,pc. They have larger velocity dispersion for a given size than clouds in the Milky Way disk and appear to represent a population of self-gravitating clouds exposed to significant external pressure. 

These massive self-gravitating structures must make up the high-$N$ end of the left panel of Figure \ref{fig4}. Meanwhile, we expect that the high \avir at lower column densities in \co32 in both galaxies likely reflect that the dendrogram picks out substructure within larger structures, and are not by themselves bound or in equilibrium.

Both panels in both galaxies show high \avir. Clouds in galaxy centers, and particularly in starbursts, are likely to have larger virial parameters than clouds in galaxy disks because of the gravity contribution from the stellar potential, the high external pressure of the environment, the widespread presence of distributed molecular material beyond bound clouds, and the substantial amount of feedback in the form of heating and turbulence.

Synthesizing, the high line widths that we observe can be partially, but not wholly, explained by self-gravity given the estimated column densities. In both galaxies, but especially NGC253, we see evidence that the higher column density structures have lower $\avir$ and appear more like self-gravitating structures than the structures with low column density. The low column-density structures show high $\avir$, indicating that the dendrogram either picks out transient out-of-equilibrium structures or equilibrium structures in which external pressure plays a dominant role in the dynamical state.

These observations fit with our picture of starburst galaxies. In a starburst the average density of the medium is such that most of the gas is molecular, and we expect a high optical depth molecular phase that is continuous and volume-filling. Analysis of NGC253 finds that the overall \co10  luminosity is dominated by a phase with low mass-to-light ratio (relative to Milky Way disk GMCs), and average column and volume densities that are large \citep[$N\sim10^{23}$~cm$^{-2}$, $n\sim300$\,\cbcm ;][]{2015ApJ...801...25L}. High resolution \co32  maps show the same picture, revealing pervasive high brightness line emission, even at few pc scales \citep{Krieger2020}. Embedded in this phase there are self-gravitating structures with masses $M(H_2)\sim3\times10^6-10^8$\,\Msun, very large surface densities ($N\sim4\times10^{23}$\,\sqcm), and large average volume densities ($n\sim2000$\,\cbcm ) \citep{2015ApJ...801...25L}. 

Finally, we note that Figure \ref{fig4} and our analysis here depends on our adopted conversion factor. We remind the reader that we adopted \alphaCO of one half the ``standard'' Milky Way value for the GC and one quarter the Milky Way value for NGC 253. The most likely deviations from this are that the low-$N$ gas actually has even lower \alphaCO, leading to even higher \avir , while the high $N$ gas shows higher \alphaCO, leading to lower \avir . That is, applying a more nuanced prescription would likely make the observed split between low and high column density gas even stronger.


\subsection{Physical Implications}
\label{section: physical implications}

The observations discussed above can be summarized in a few key points: 1) the velocity dispersion on any given size scale in the range $1{-}100$\,pc is about $2.5$ times larger in NGC253 than in the GC. 2) The relation between luminosities and sizes is similar for NGC253 and the GC in \co10 on $8{-}50$\,pc scales, although in \co32  NGC253 is a factor of $\sim3$ more luminous than the GC on scales of $1{-}10$\,pc. 3) Most structures on scales of $1-10$\,pc are either pressure bound or transient in both NGC253 and the GC. And, 4) on scales of $10{-}100$\,pc structures in the GC appear mostly compatible with self-gravity equilibrium, while in NGC253 this is only true for column densities over $N\sim3\times10^{22}$\,\sqcm (equivalent to $\Sigma_{mol} \sim 500$\,\Msun\,pc$^{-2}$).

These suggest that there is a widespread, highly turbulent molecular medium in the NGC253 starburst,  with higher excitation than the GC on small scales. The latter is simply a consequence of the starburst activity and is well-supported by observations. For example, we know that CO excitation peaks around the $J=7-6$ transition in NGC253 and near the $J=4-3$ or $5-4$ transitions in the center of the Milky Way \citep{bennett1994,bradford2003}. The strong departure from self-gravity for column densities lower than $N\sim3\times10^{22}$~cm$^{-2}$, with essentially constant $\sigma^2/R\sim7-15$~km$^2$\,s$^{-2}$\,pc$^{-1}$ depending on the tracer, suggests either very high pressures $P/k\sim10^6-10^7$~K\,cm$^{-3}$ or a medium with $\alpha_{vir}\sim10-100$. Although the bulk gas temperatures in NGC253 are $T\sim50{-}100$\,K \citep{bradford2003,2015ApJ...801...63M}, larger than averages in the GC, the implied pressures are much in excess of plausible average thermal pressures in the system. Therefore most of CO emission for $N\lesssim3\times10^{22}$~cm$^{-2}$ is not tracing bound, equilibrium structures in NGC253. It must correspond to a widespread, volume filling molecular phase with an enhanced level of turbulence fed by the starburst activity. This phase does not have a clear correspondence in the GC.


\section{Summary and conclusion}
\label{section: summary}

We perform a resolution-, area- and noise-matched comparison of molecular cloud properties in the starburst center of NGC253 and the Milky Way Galactic Center. We compare ALMA observations of NGC253 in \co10 and \co32 to data for the Galactic Center from the COGAL and CHIMPS2 surveys. Using \astrodendro, we decompose the structure of the observed emission and compare the respective size -- line width, size -- luminosity, and $\sigma^2/R$ -- column density relation (related to the virial state and external pressure) over a matched range of spatial scales (approximately $R\sim1{-}10$\,pc and $R\sim10{-}100$\,pc for \co32 and \co10 respectively).

In the following, we briefly summarize our work and present our conclusions.

\begin{enumerate}[noitemsep,topsep=0pt]
    
\item The size -- line width relations in NGC253 and the GC show comparable slopes, $0.7{-}0.8$, but at any given size scale the velocity dispersion is larger in NGC253 than the GC by a factor of $\sim2.5$.

\item NGC253 and the GC follow similar size -- luminosity relations with $L \propto R^3$, suggesting roughly constant volume density in the dendrogram-selected structures over the explored range of size scales.

\item The $\sigma^2/R$ -- column density relation shows that the increased line widths in NGC253 originate in low column density gas ($N \lesssim 3 \times 10^{22}$\,\sqcm) gas while at high column density ($N \gtrsim 3 \times 10^{22}$\,\sqcm) NGC253 and the GC occupy similar parameter space.

\item On the $R\sim10{-}100$\,pc scales sampled by the \co10 emission, structures in the GC with column densities over $N\sim3\times10^{21}$\,\sqcm show typical $\alpha_{vir}\sim2-3$ compatible with equilibrium under self-gravity. Structures in NGC253 are only compatible with equilibrium under self-gravity for $N\gtrsim3\times10^{23}$\,\sqcm. At lower column densities and/or on smaller spatial scales, most structures have higher $\avir$. This implies that the low-$N$ structures are either transient or possibly in pressure-bound virial equilibrium. Given that the bounding pressures appear implausibly high for $N\gtrsim10^{22}$\,\sqcm , we prefer the explanation that the dendrograms pick out transient structures at these size scales and column densities.

\item The decoupling between $\sigma^2/R$ and column density observed in NGC253 below $N\sim10^{23}$\,\sqcm is likely due to a widespread molecular phase that is not bound in clouds, but that likely fills most of the volume. Such a volume-filling phase is already suggested by the pervasive high brightness emission seen in \co32  maps of NGC253 \citep{2019ApJ...881...43K}. The excess kinetic energy of the unbound molecular gas in NGC253 relative to the GC is most plausibly supplied by feedback from the starburst. 
\end{enumerate}

\acknowledgements
The authors would like to thank the anonymous referee for helpful comments on this work.
The authors further thank Harriet Parsons for valuable feedback on the draft of this paper.
This paper makes use of the following ALMA data: ADS/JAO.ALMA \#2011.1.00172.S and \#2015.1.00274.S. ALMA is a partnership of ESO (representing its member states), NSF (USA) and NINS (Japan), together with NRC (Canada), NSC and ASIAA (Taiwan), and KASI (Republic of Korea), in cooperation with the Republic of Chile. The Joint ALMA Observatory is operated by ESO, AUI/NRAO and NAOJ.

The CHIMPS2 data are part of the JCMT Large Programs with data obtained under project code M17BL004.  The James Clerk Maxwell Telescope is operated by the East Asian Observatory on behalf of The National Astronomical Observatory of Japan; Academia Sinica Institute of Astronomy and Astrophysics; the Korea Astronomy and Space Science Institute; Center for Astronomical Mega-Science (as well as the National Key R\&D Program of China with No. 2017YFA0402700). Additional funding support is provided by the Science and Technology Facilities Council of the United Kingdom and participating universities in the United Kingdom and Canada.

EK and ER acknowledge the support of the Natural Sciences and Engineering Research Council of Canada (NSERC), funding reference number RGPIN-2017-03987. 
The work of A.K.L. is partially supported by the National Science Foundation (NSF) under Grants No.1615105, 1615109, and 1653300, as well as by the National Aeronautics and Space Administration (NASA) under ADAP grants NNX16AF48G and NNX17AF39G. EACM gratefully acknowledges support by the National Science Foundation under grant No. AST-1813765.

\facilities{ALMA, JCMT, CfA 1.2\,m Millimeter-Wave Telescope}
\software{Astropy \citep{Collaboration:2013cd,Collaboration:2018ji}, Astrodendro\footnote{dendrograms.readthedocs.io}, NumPy \citep{van2011numpy}, SciPy \citep{scipy}, spectral-cube\footnote{https://spectral-cube.readthedocs.io}}, CASA \citep{McMullin:2007tj}


\clearpage
\appendix


\section{Definition of structure properties}
\label{appendix: structure definition}

The exact definition of size and line width of a structure can influence the derived scaling relations and inferred physical state of the gas \citep[e.g.][]{2002ApJ...570..734B,2010ApJ...712.1049S}. In this section, we explore the effect of different size and line width definitions on derived properties such as the size -- line width relation.


\subsection{Structure size}
\label{appendix: size definition}

The projected two-dimensional size of a structure can either be defined by its linear extent in some direction(s) or via the covered area. Using the area takes the often complex shape of structures into account, but does not account for the distribution of gas within a structure. In an extreme case, 99\% of the mass might be inside 1\% of the area and a good definition of structure size should come up with a size much smaller then the total extend of the cloud.

By default, \astrodendro (\texttt{radius}) defines ``size'' as the mean structure radius of the intensity-weighted second moment map. Mean structure radius is defined as the mean of major axis in the direction of greatest elongation and the minor axis perpendicular to the major axis. In this comparison, we denote this definition as R$_\mathrm{astrodendro}$.

We compare R$_\mathrm{astrodendro}$ to two other definitions. First, the mean radius for an ellipse fitted to the structure, R$_\mathrm{ellipse}$, and the mean radius of a circle with area equal to the projected structure, R$_\mathrm{circular} = \sqrt{\mathrm{A}/\pi}$, where $A$ is the projected area of a structure.

\begin{figure*}
    \centering
    \includegraphics[width=\linewidth]{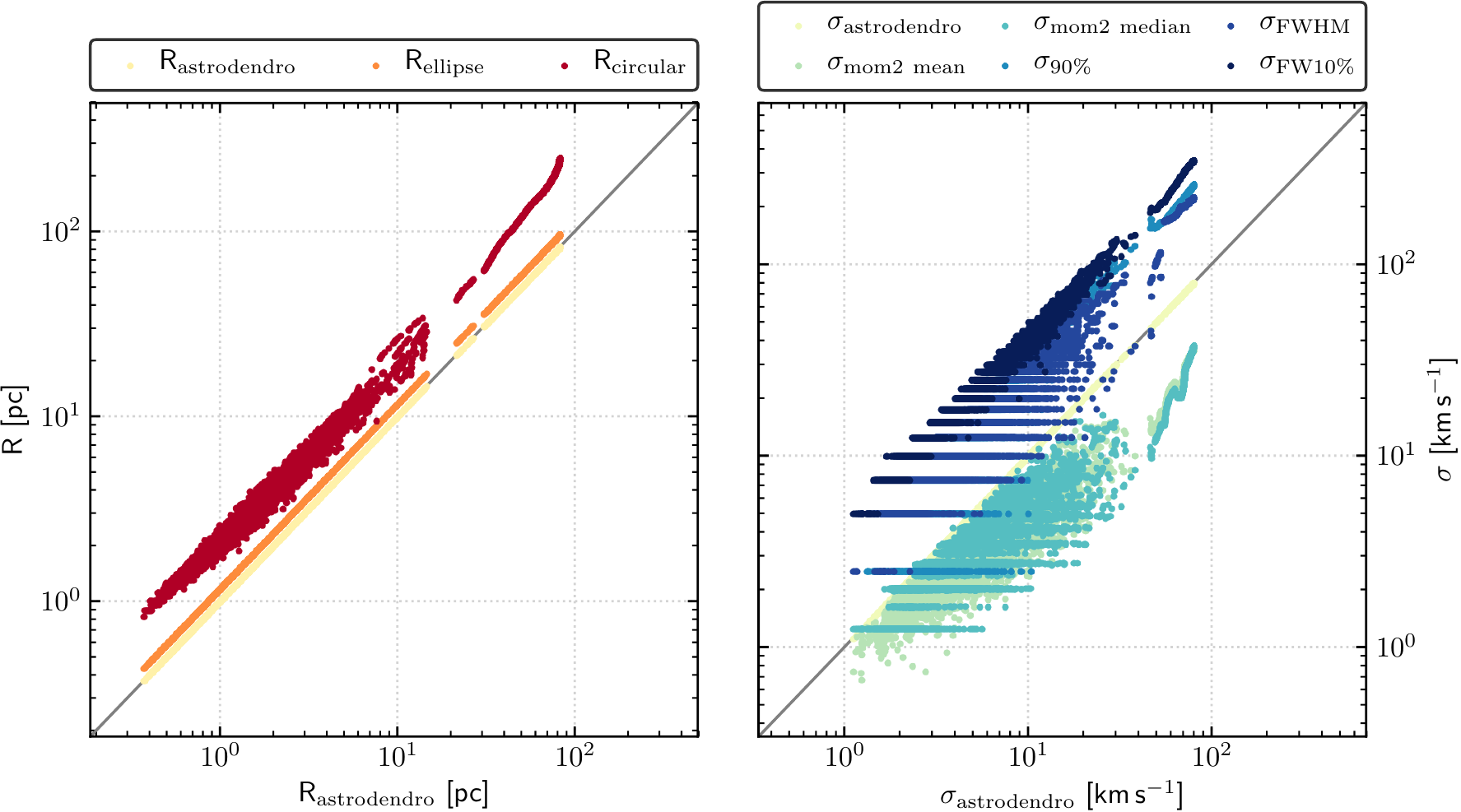}
    \caption{Comparison of different definitions for structure size (\emph{left}) and line width (\emph{right}). 
    The suffix astrodendro refers to the quantities calculated by \astrodendro directly as described in Section~\ref{section: dendrogram}.
    For details on the other definitions for size and line width see Appendix~\ref{appendix: size definition} and \ref{appendix: line width definition}, respectively.
    Quantization apparent in some line width definitions follows directly from the discrete channel structure of the data cubes.
    }
    \label{figA}
\end{figure*}

In Figure~\ref{figA} (left), we compare these three size estimates for \co32 in NGC253. The moment-based R$_\mathrm{astrodendro}$ and R$_\mathrm{ellipse}$ track one another almost perfectly, modulo a small but fixed multiplicative offset related to their definitions. R$_\mathrm{circular}$ yields size a factor $\sim 2$ larger because it reflects the overall footprint of the structure and not the intensity distribution. Over more than two orders of magnitude R$_\mathrm{circular}$ remains almost parallel to the other size definitions. 

We use R$_\mathrm{astrodendro}$, but this comparison shows that had we selected one of the other size definitions, the main effect would be to shift the normalization of the sizes.

\begin{figure}
    \centering
    \includegraphics[width=0.5\linewidth]{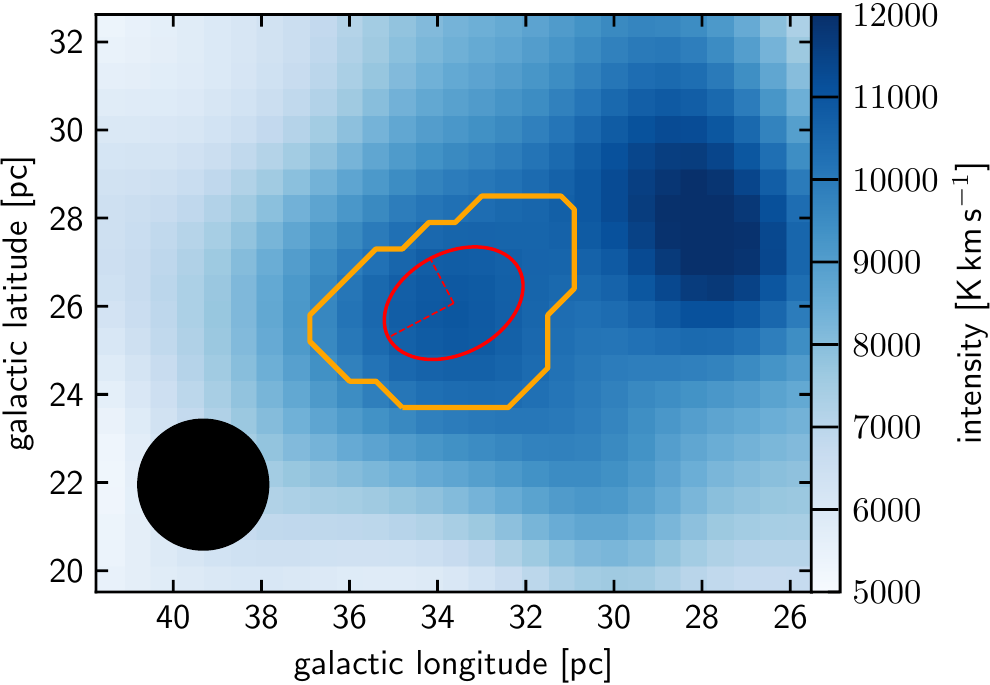}
    \caption{Example how \astrodendro assigns sizes to structures. The figure shows a random structure (\#4822) in the \co32 dataset of NGC253 which describes a local emission peak. The total extent of the structure (orange) and the corresponding intensity-weighted second moment ellipse (red) are plotted on top of the total integrated intensity (0$^\mathrm{th}$ moment) \co32 map (blue). Dashed lines indicate the semi-major and semi-minor axes whereof the mean defines the size of the structure, here R = 1.20\,pc. The beam of 3\,pc FWHM is shown in the bottom left corner.
    Note that the background image shows all data to provide context whereas the size ellipse is calculated for the particular structure within the orange line.
    In this special case of a small structure, the \astrodendro algorithm assigns a size smaller than the beam to the structure. Nevertheless, the structure is resolved.
    \label{fig1}}
\end{figure}

Note that \astrodendro parametrizes size as a semi axis (instead of full axis) which allows for resolved structures apparently smaller than the spatial resolution (given as FWHM). Furthermore, this definition incorporates intensity weighting and will thus assign sizes smaller than half the resolution to structures approaching to the resolution limit. The exact value depends on the distribution of emission within the structure. The minimum PPV volume threshold we chose for this analysis ensures that a structure is resolved and derived sizes smaller than the resolution do \emph{not} mean that a structure is unresolved.
Figure~\ref{fig1} shows an example of a small but resolved structure in \co32 in NGC253. The size inferred by the \astrodendro algorithm is 1.20\,pc and thus less than half the FWHM beam size.


\subsection{Structure line width}
\label{appendix: line width definition}

Similarly to size, the line width can be defined in various ways, which have different response to the spectral and spatial distribution of emission within a structure.

In our analysis, we use the default \texttt{v\_rms} quantity in \astrodendro , which we label $\sigma_\mathrm{astrodendro}$ here. This quantity represents the second moment of the integrated spectrum of the structure in question. In the right panel of Figure \ref{figA}, we compare this to four other line width metrics. First, $\sigma_\mathrm{mom2\ mean}$ and $\sigma_\mathrm{mom2\ median}$ represent the median and mean of the second moment map over the footprint of the structure. These measurements effectively remove scatter in the mean velocity from line-of-sight to line-of-sight. 
We also show results for $\sigma_{90}$, the line width that captures 90\% of the emission in the integrated spectrum, $\sigma_{\rm FWHM}$, the full width at half-maximum of the integrated spectrum, and $\sigma_{\rm 10}$, the width at 10\% of the maximum for the integrated spectrum. Similar to the use of area above, these other line width measures are sensitive to the shape of the spectrum in different ways than the second image moment.

Note that we do not correct all of our line width measures onto a common system. That is, $\sigma_{FWHM}$ is the full width at half maximum of the line. Even for an ideal Gaussian line we expect it to differ from the rms line width by a factor of $2.354$. This leads to systematic offsets in Figure~\ref{figA} without any actual difference in measured line width.

Figure~\ref{figA} (right) shows the comparison of these six line width definitions for \co32 in NGC253. Aside from the highest line width structures, which represent the trunks and lowest branches in the dendrogram tree, the different definitions lie approximately parallel to each other indicating that only the normalization changes. Choosing a different definition of line width will thus not distort derived relations but merely shift them.


\section{Size -- mass relation}
\label{appendix: size mass}

\begin{figure*}
    \centering
    \includegraphics[width=\textwidth]{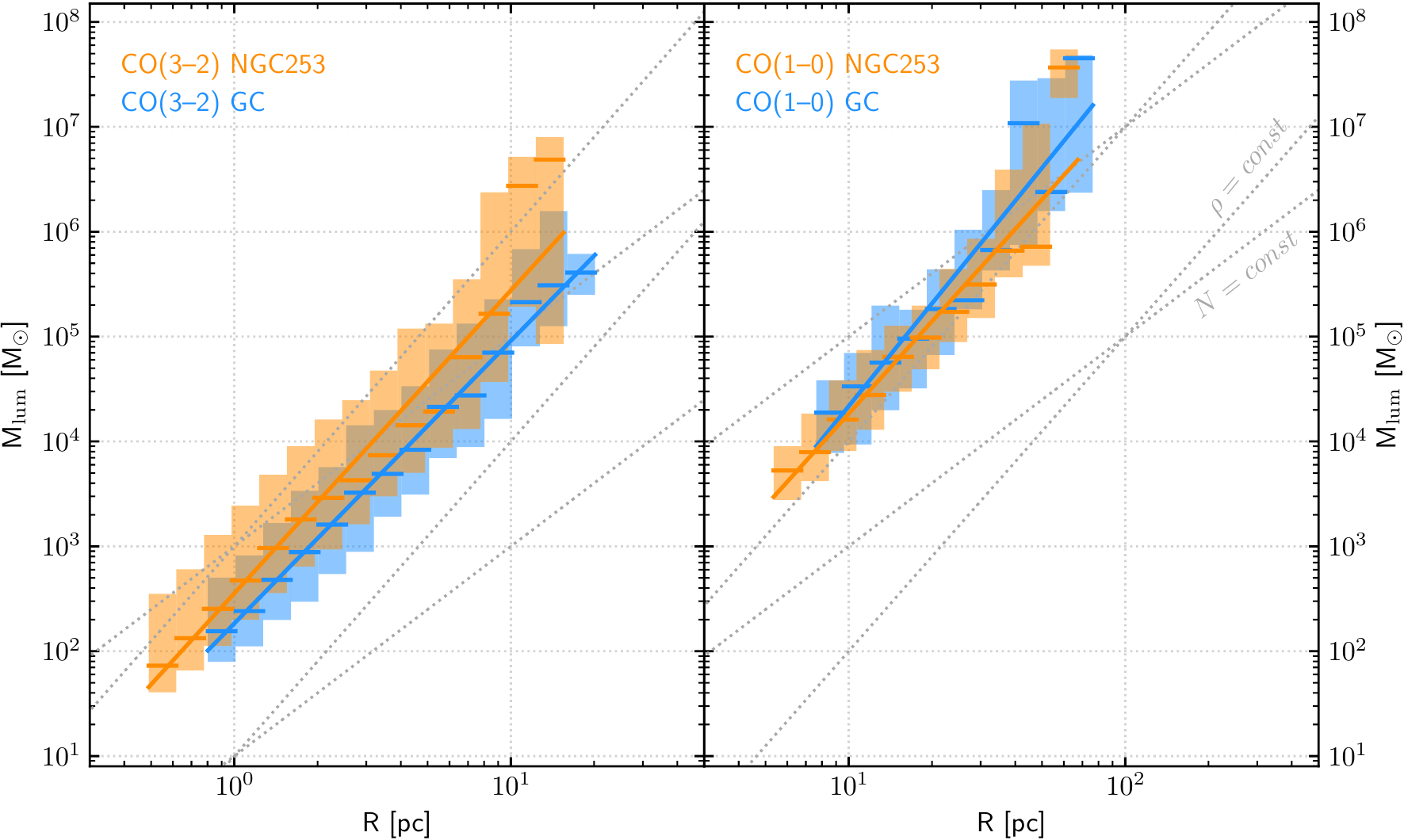}
    \caption{Relation between dendrogram structure size and mass for \co32 (\emph{left}) and \co10 (\emph{right}) in NGC253 and the GC. Horizontal lines indicate the median of the distribution of masses (colored bars) in each bin. The power law fits (solid lines) correspond to those to the size -- luminosity relation (Table~\ref{table3}). The masses are derived applying our adopted conversion factors, with \alphaCO for NGC253 two times lower than for the GC. In each panel, two lines of constant surface density $N$ ($M \propto R^2$) and constant volume density $\rho$ ($M \propto R^3$) are shown for reference.
    \label{figB}}
\end{figure*}

In Figure~\ref{figB}, we show the size--mass relation. This figure is derived from Figure~\ref{fig3} by applying our adopted CO-to-H$_2$ conversion factors. Because we adopt \alphaCO two times smaller in NGC253 compared to the GC, the mass-radius relations appear more similar than the size -- luminosity relations. That is, NGC253 is more luminous than the GC in \co32, but our adopted conversion factor removes this difference from the mass-based relation.

\section{Virial state of the gas separated by size scale}
\label{appendix: virial state separated}

\begin{figure*}
    \centering
    \includegraphics[height=0.9\textheight]{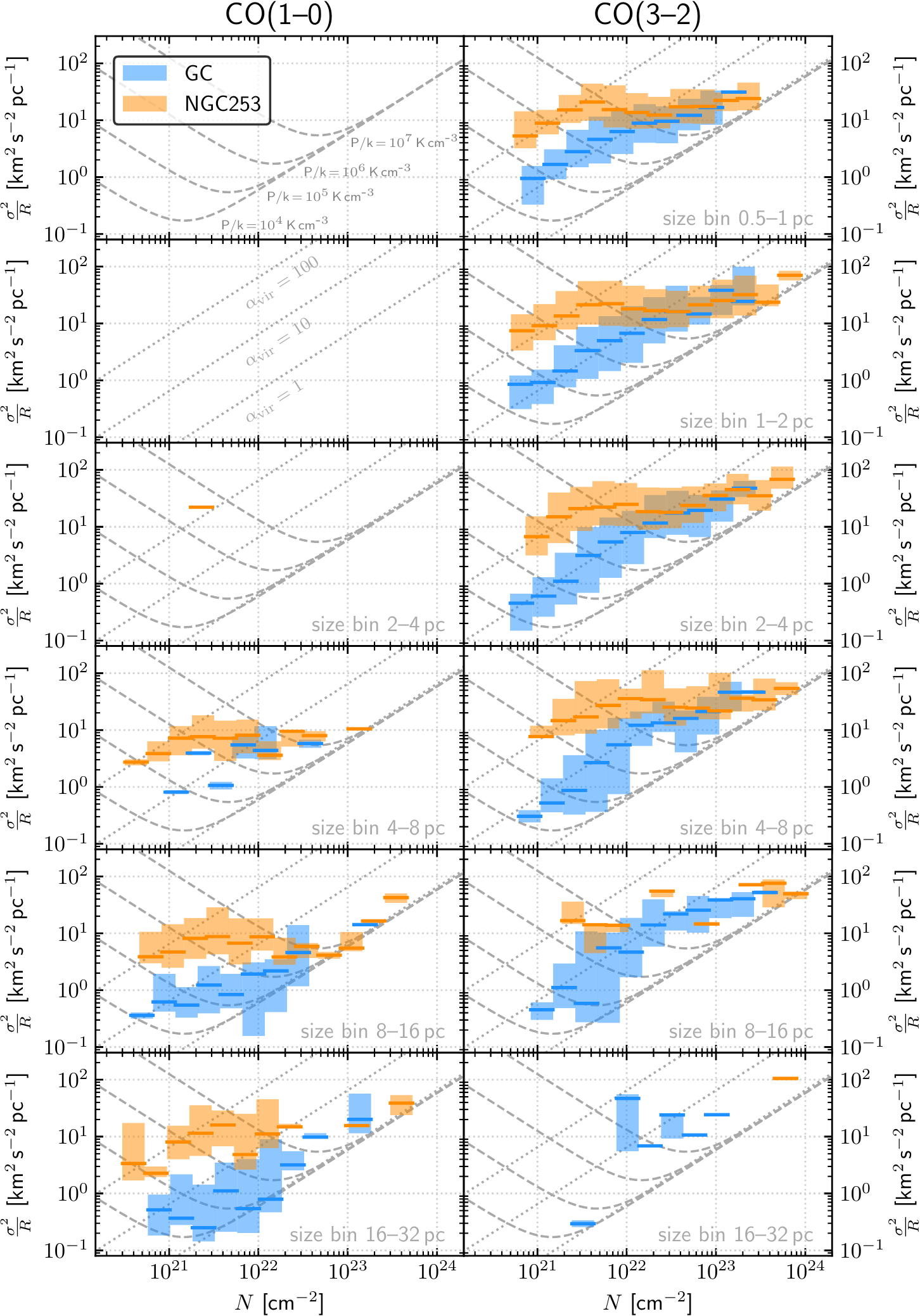}
    \caption{Size--line width coefficient $\sigma^2/R$ as a function of column density for a range of structure size bins. Diagonal lines indicate lines of constant virial parameter under the assumption of idealized spherical clouds (cf. Section~\ref{section: virial state}). Dashed lines represent lines of constant external pressure on a spherical cloud (cf. Section~\ref{section: virial state}). Horizontal lines indicate the median of the distribution of $\sigma^2/R$ (colored bars) in each bin.\\
    \emph{Left}: low resolution \co10; \emph{Right}: high resolution \co32. The size bin for each panel is given in the lower right corner. 
    \label{figD}}
\end{figure*}

The interpretation of Figure~\ref{fig4} is complicated by the fact that the size -- line width coefficient $\sigma^2/R$ is degenerate between $\sigma$ and $R$. In Figure~\ref{figD}, we keep the size fixed (up to a factor of two), so that any change in $\sigma^2/R$ must be driven by $\sigma$.

Aside from the fact that there are very few bins $<4$\,pc in \co10 and $>16$\,pc in \co32, there is no relevant deviation between the six vertical panels of Figure~\ref{figD}. The collapsed data as shown in Figure~\ref{fig4} thus captures the complete picture.

\clearpage
\bibliography{bibliography.bib}


\end{document}